\documentclass{article}

\usepackage{arxiv}

\usepackage[utf8]{inputenc} 
\usepackage[T1]{fontenc}    
\usepackage{hyperref}       
\usepackage{booktabs}       
\usepackage{amsfonts}       
\usepackage{nicefrac}       
\usepackage{microtype}      
\usepackage{cleveref}       
\usepackage{lipsum}         
\usepackage{graphicx}
\usepackage{natbib}
\usepackage{doi}

\usepackage{xurl}
\usepackage{tikz}
\usetikzlibrary{positioning, arrows.meta, decorations.pathmorphing}
\usepackage{verbatim}
\newsavebox{\mybox} 

\usepackage{subcaption}

\title{Onions in the ATmosphere: Decentralizing Trust in Identity}

\date{September 9, 02026\thanks{We follow herein date
conventions of the Long Now Foundation, {\small
  \texttt{https://www.longnow.org/ }}.}}

\newif\ifuniqueAffiliation

\ifuniqueAffiliation 
\author{ \href{https://orcid.org/0000-0000-0000-0000}{\includegraphics[scale=0.06]{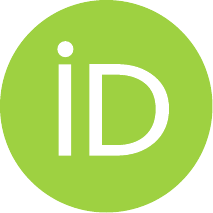}\hspace{1mm}David S.~Hippocampus}\thanks{Use footnote for providing further
		information about author (webpage, alternative
		address)---\emph{not} for acknowledging funding agencies.} \\
	Department of Computer Science\\
	Cranberry-Lemon University\\
	Pittsburgh, PA 15213 \\
	\texttt{hippo@cs.cranberry-lemon.edu} \\
	\And
	\href{https://orcid.org/0000-0000-0000-0000}{\includegraphics[scale=0.06]{orcid.pdf}\hspace{1mm}Elias D.~Striatum} \\
	Department of Electrical Engineering\\
	Mount-Sheikh University\\
	Santa Narimana, Levand \\
	\texttt{stariate@ee.mount-sheikh.edu} \\
}
\else
\usepackage{authblk}

\newbox{\orcid}\sbox{\orcid}{\includegraphics[scale=0.06]{orcid.pdf}}
\author[1]{%
	\href{https://orcid.org/0000-0001-6887-3658}{\usebox{\orcid}\hspace{1mm}Paul Syverson\thanks{\texttt{paul.f.syverson.civ@us.navy.mil}}}%
}
\author[2]{%
	\href{https://orcid.org/0009-0000-4874-615X}{\usebox{\orcid}\hspace{1mm}Nick Gerakines\thanks{\texttt{nick.gerakines@gmail.com}}}%
}
\affil[1]{Center for High Assurance Computer Systems, U.S. Naval Research Laboratory, Washington, DC 20375, USA}
\affil[2]{Independent, USA}
\fi

\renewcommand{\headeright}{}
\renewcommand{\undertitle}{}
\renewcommand{\shorttitle}{Onions in the ATmosphere}

\hypersetup{
pdftitle={Onions in the ATmosphere: Decentralizing Trust in Identity},
pdfauthor={Paul Syverson, Nick Gerakines},
pdfkeywords={onion service, decentralized identifier, AT Protocol, identity, Bluesky trusted verifier, blocking resistance, manipulation resistance, contextual trust},
}

\begin{document}
\maketitle


\begin{abstract}
  Multiple approaches have been implemented for identities that
  associate meaningful identifiers, such as domain names, with
  self-authenticating identifiers, such as onion addresses. While
  these approaches offer various advantages both from identifier
  meaningfulness and from identifier self-authentication, identifier
  association itself is subject to blocking, hijack, fingerprinting,
  and other forms of surveillance and manipulation. So too are the
  resources associated with an identity comprising a meaningful
  identifier and a self-authenticating identifier.  This paper
  describes ongoing work on two such approaches to self-authenticating
  meaningful identity (SAMI): associating a registered domain name
  with an onion address (onion association) and associating an atproto
  handle with a decentralized identifier (DID association).  We
  present both the application of already implemented and published
  means of secure onion association to DID association and the
  combination of both in a single identity. Finally, we describe
  grounding identity association and integrity in contextual trust,
  wherein trusting attestations of identity relies on contextual
  relevance of an attestor to the identity rather than relying only on
  structural authority, for example as embodied in the traditional
  centralized infrastructure.
\end{abstract}




\section{Introduction}
\label{sec:intro}

Identifiers such as domain names are generally meaningful to humans
and infrastructure (e.g., DNS), but they require authentication (e.g.,
as provided by TLS) to be securely usable. Using the associated
authentication infrastructure is itself subject to various
manipulations and attacks.  Alternatively, self-authenticating
identifiers (e.g., those that encode a public key) are generally not
subject to the same attacks, but they are also typically not
meaningful.  For this reason, a self-authentication meaningful
identity (SAMI), which comprises meaningful and self-authenticating
identifiers, is useful.

Two approaches to SAMI are (1)~the \emph{onion
association} of a web address (domain name) with a
(self-authenticating) onion address, and (2)~the \emph{DID
association} of an atproto (Authenticated Transfer Protocol) identity
comprising a (meaningful) \emph{handle} and a (self-authenticating)
\emph{DID} (Decentralized Identifier). The AT Protocol actually
recognizes two types of DIDs, only one of which is
self-authenticating. We will briefly describe the two types and other
basic background on these two SAMI approaches below in
Sections~\ref{sec:basic-background-onion-association}
and~\ref{sec:atproto-background} respectively. Because an atproto
handle can be a registered domain name, and an onion address can be
associated with a DID recognized by the AT Protocol, a SAMI can
effectively be both a DID association and an onion association.  (We
will follow general practice herein, using ``AT
Protocol'' and ``atproto'' interchangeably.)

Associating meaningful and self-authenticating identifiers presents
new problems: how to discover the associated self-authenticating
identifier given the meaningful identifier (or vice versa) and how to
guarantee that the discovered association is correct. Establishing
trust in such association is traditionally structural (and
mostly centralized): e.g., Certificate
Authorities (CAs) are generally trusted system-wide to issue
TLS certificates attesting that a public key is associated with a given
domain name. And a CA will issue a certificate to anyone satisfying
standardized checks that they meet specific criteria.

Traditional identifier discovery and identity integrity are subject to
blocking, hijack, lack of transparency, targeted surveillance, and
other manipulations---so too are discovery of and access to
identity-associated resources and content. An alternative to relying only
on the traditional trust infrastructure is to also
ground trust in contextually relevant attestors. So a particular
company, consortium, government, professional organization,
hobby or interest group, etc.\ can be trusted for attestations for all
and only identifier associations for members or affiliates.
Separately, each of the two noted approaches to SAMI has already
implemented or proposed countermeasures to some of these problems. 
A primary goal of this paper is to present techniques that
address all of them, doing so in a way that leverages some of
the positive features of each.

Another goal is to address these problems within existing protocols
and infrastructure. We avoid mechanisms that require changes to
existing standards or current practices of widely used infrastructure
elements on which they depend. For example, though issuance of TLS
certificates for onion addresses within ACME is an IETF Draft Standard
as of June 02025~\cite{acme-onions-rfc}, at time of writing we do not
know when the automated issuance of certificates by CAs such as Let's
Encrypt will be supported. 
Our mechanisms make use of existing certificate issuance practices of
Let's Encrypt and others. And though in the long run we hope that our
approach is fully integrated into widely used browsers and other
software by default, for now the client end is reflected in a browser
extension. Some of our contributions do not even require that. For
example, nothing other than configuration of existing
broadly-available systems is needed for basic authenticated,
transparent discovery of and access to a website from a meaningful
identifier \emph{without} requiring a registered
domain~\ref{sec:handles-for-website-names}.

Relatedly our solutions are backwards compatible with widely used
browsers. Those not configured to utilize the discovery and protection
mechanisms of our URLs and headers will not benefit from them, but
they will respond the same as when interacting with the same host if
our mechanisms were not present. Our solutions requiring access to
onion services generally require a browser configured to access the
Tor network, such as Tor Browser or Brave. Some of our contributions,
however, do not require onion services or Tor network access at all.

\subsection{Contributions}
\label{sec:contributions}

Contributions set out in this paper include:

\begin{itemize}

  \item techniques to synthesize the two types of DIDs used by atproto
    to improve manipulation resistance and enhance transparency
    (Sec.~\ref{sec:synthesizing-did-methods}),
  
  \item application of implemented and published means
    of securing onion association to atproto identities comprising DID
    associations (Sec.~\ref{sec:synthesizing-did-methods}),

  \item the first tool to support checking validation of the standard
    atproto requirement of bidirectional linking between handles and
    DIDs (Secs.~\ref{sec:handle-hijack},~\ref{sec:implementation}),
    
  \item single identities that can simultaneously include
    identifiers for both of these two approaches to SAMI and thus
    broader, more resilient means of protection as well as wider
    access to blocking-resistant and manipulation-resistant systems
    than either existing onion association mechanisms or existing
    atproto mechanisms currently offer on their own (Sec.~\ref{sec:did-onion}),
    
  \item illustration of the above in an implementation for and
    deployment on the popular atproto event management app, Smoke Signal
    (Sec.~\ref{sec:did-onion}),

  \item a practical technique based on our SAMI mechanisms to use
    atproto handles as names for personal websites without having to
    register a domain name and that is easily deployed using existing
    systems (Sec.~\ref{sec:handles-for-website-names}),
    
  \item describing contextual trust and how it can by used to
    encompass Bluesky's verified accounts and provide identity trust
    that is more generally available and more decentralized
    (Sec.~\ref{sec:contextual-trust}),

  \item design and partial implementation of a current complement to
    and potential successor of Bluesky trusted verifiers that is more
    manipulation resistant and closer to the atproto goal of autonomy
    over one's identity and data
    (Secs.~\ref{sec:contextual-trust-for-trusted-verifiers},~\ref{sec:integation-with-atproto}),

  \item implementation of the above techniques and functionalities,
    available in FLOSS client and server software, linked to and
    glossed in the text (Sec.~\ref{sec:implementation}),

  \item a way to make onion discovery and onion association more
    secure, as well as doing so at scale via integration with atproto
    identities, an existing large-scale deployed system of
    decentralized identities (Secs.~\ref{sec:did-onion},~\ref{sec:contextual-trust}).

\end{itemize}

\subsection{Basic background: AT Protocol identity}
\label{sec:atproto-background}

The atproto ecosystem (aka the ATmosphere) uses DIDs, a W3C
standard for decentralized identifiers~\cite{did-w3c-standard}. Of the
many dozens of recognized DID methods, atproto currently accepts two.

The {\small \texttt{did:web}} method uses a domain name for the DID
and stores the DID document (which lists public keys, accepted
authentication protocols, service endpoints, etc.) in a JSON file in a
standard location on the webpage. As an example, for the event
management and community building site Smoke Signal the did-method-web
DID is {\small \texttt{did:web:smokesignal.events}} and the
corresponding DID doc is found at
{\small \texttt{https://smokesignal.events/.well-known/did.json}}.  This
method allows the atproto handle {\small \texttt{@smokesignal.events}}
to effectively be the same as the DID\@. The preferred method of
handle resolution for a domain-name handle is when ``a DNS TXT record
is registered for the {\small \texttt{\_atproto}} sub-domain under the handle
hostname''~\cite{atproto-spec-handle}.  Unsurprisingly, by themselves
handle resolution by this means and/or accessing the DID doc when
hosted as just described are vulnerable to all of the well-known
blocking and manipulation problems to which traditional web domains
are generally vulnerable.
(Cf.~\cite{domain-impersonation-acmccs2019,onion-discovery-attacks}
for further citations and for a sampling of original examples of such
blocking and manipulation.)

The {\small \texttt{did:plc}} method was originally devised by Bluesky
for use with its Public Ledger of Credentials (PLC).  Handles can be
almost any unique string, which then can correspond to a subdomain of
{\small \texttt{bsky.social}}, but a handle can also be a domain name
such as {\small \texttt{@smokesignal.events}} or {\small
\texttt{@torproject.org}} (in
which case {\small \texttt{bsky.app}} does not treat it as a subdomain of
{\small \texttt{bsky.social}}).  A {\small \texttt{did:plc}}
identifier is a string comprised of twenty-four base-32 characters
generated from a hash of the \emph{genesis}, the first operation for
that DID in the transparent log at the heart of the PLC\@. The
did-method-plc DID for {\small \texttt{@torproject.org}} is {\small
  \texttt{did:plc:tnyyuxpjrcuf2sispj2m7bdc}}.  Thus, unlike when using
the {\small \texttt{did:web}} method, a did-method-plc DID is a
meaningless string without obvious connection to the handle.  But it
is self-authenticating, and it, along with updates to its DID doc, is
recorded in a transparent, publicly verifiable log.  Recall that a
SAMI comprises a meaningful identifier and a
\emph{self-authenticating} identifier. Likewise \emph{DID association}
is association between a meaningful handle and a
\emph{self-authenticating} DID\@. Thus an association between a handle
and a did-method-web DID is not a DID association in our usage and is
not sufficient to form a SAMI\@.

In the ATmosphere, both a did-method-web DID and a
did-method-plc DID are considered an ``identity''. For our purposes herein,
however, an identity is a SAMI, i.e., an entity comprised of both
a self-authenticating identifier and a meaningful identifier. Thus for
us, neither a did-method-web DID nor a did-method-plc DID is by itself
an identity. We refer to each of those as identifiers.

Extensive systems for identity management are only of value if
identities do something. Content produced by an atproto identity is
served from a \emph{repository} at a \emph{Personal Data Server} (PDS)
which is publicly available as a web service and identified in the DID
doc.  A primary component of the atproto design is the
\emph{relay}, which crawls all repositories at all PDSes for record
updates. This is output in the \emph{firehose}, an aggregate stream
that is consumed by feed generators and other atproto network
components. An \emph{AppView} gathers records from the firehose as
well as media files from PDSes and processes these for a webservice it
provides to clients (along with a few other functions).
AT protocol records are referenced via the IANA registered atproto URI
scheme, {\small \texttt{at://}}, which can be followed by either a handle or a
DID~\cite{atproto-spec-at-uri-scheme}.
For a more detailed intro to atproto, its DID methods, and Bluesky than
is offered in this section, see~\cite{bluesky-paper}.

\subsection{Basic background: onion association}
\label{sec:basic-background-onion-association}

Onion addresses are IETF standard domain names, not resolvable in
DNS~\cite{ietf-onion-tld-rfc}. They are 56-character strings that
encode their own public key. Thus, like did-method-plc DIDs, they are
self-authenticating.  Among their other properties, their onion
service protocols also make integral use of the anonymity network
Tor~\cite{tor-design}, providing secure and private address lookup and
routing. Address lookup uses a directory system comprised of Tor nodes
called \emph{HSDirs} that are arranged in a DHT (distributed hash table).
Onion addresses also provide {\small \texttt{.onion}} domain owners greater
autonomy over site access and authentication when compared to
traditional domain names.  For a more detailed introduction to onion
addresses, onion services, their properties, how to set them up,
existing applications, and featured examples of onionsites
see~\cite{onion-service-overview,onion-service-ecosystem}.

Multiple means of onion association have been proposed and
deployed. Of these, Onion-Location
(O-L)~\cite{onion-location-overview} is probably the most well-known
and widely used~\cite{olf-popets}).  O-L offers a radio button in the
URL bar labeled ``.onion available'' when connected via Tor Browser to
supporting DNS domains.  When this button is clicked, the client looks
up the onion address in the DHT and connects to the corresponding
onion service.  Brave Browser also offers a similar feature.

\subsection{Related Work}
\label{sec:related-work}

Multiple descriptions of blocking, hijack, fingerprinting, and other
attacks on Onion-Location and other means of onion association have been
published, as have alternative solutions that leverage TLS
certificates~\cite{secdev19,onion-discovery-attacks} and the
Certificate Transparency (CT) infrastructure.  Like the PLC, CT logs
are public, append-only logs. They facilitate transparency in TLS
certificate issuance. Onion association that uses CT logs for
discovery as well as for transparency is called~\emph{sauteed
onions}~\cite{sauteed-onions-wpes,olf-popets}.

Sauteed onions use SANs (Subject Alternative Names) in TLS
certificates comprised of onion addresses as subdomains of registered
domain names~\cite{once-and-future}.  Such onion-association SANs are
generally used to support ultimate connection to an onion service (via
onion service protocols that utilize the Tor network).
The current TLS certificate for {\small \texttt{mullvad.net}} illustrates this
approach: one of the SANs in that certificate is
{\small \texttt{o54hon2e2vj6c7m3aqqu6uyece65by3vgoxxhlqlsvkmacw6a7m7kiadonion.mullvad.net}}. Because it is comprised of a meaningful
identifier and a self-authenticating identifier, such a SAN is a
form of SAMI\@. The association is only in the certificate, thus
authenticated by the CA based on validation during certificate issuance.
Association is not yet confirmed by the self-authenticating onion address.

Self-Authenticating Traditional Addresses (SATAs) are URLs including
both a domain name and an onion address. They also use a certificate
with such a SAN as well as a header signed by the onion address
authenticating that the association matches the TLS
certificate~\cite{onion-discovery-attacks}. Among other things, this
implies that the self-authenticating identifier can authenticate the
association when connecting to the clearnet URL; so using onion service
protocols on the Tor network is supported but optional for SATAs.
Thus, with SATAs it is
also possible to use the onion address only for the additional
authentication guarantees and censorship-resistant association
lookup. The SATA WebExtensions described in~\cite{secdev19}
and~\cite{onion-discovery-attacks} were thus useful to run in Firefox,
not just Tor Browser. SATAs also support contextual trust labels, as
discussed in Sec.~\ref{sec:contextual-trust}.
The SAN encoding of onion association in the certificate is, at the
time of writing, the same regardless of whether for use with sauteed onions or
full SATAs. There is, however, an active Tor Proposal to specify distinct
encodings for the distinct cases~\cite{oasic-tor-proposal}.

Bluesky's trusted verifiers, about which we will say more in the next
section and in Sec.~\ref{sec:contextual-trust}, are a type of
contextually relevant attestor inasmuch as they attest to the identity
of Bluesky accounts affiliated with
them~\cite{bluesky-trusted-verifier-announced}.  Such trusted accounts
display a ``verification badge'' on the Bluesky app. These badges are
the same regardless of which trusted verifier issued them, though
clicking on the badge will reveal which trusted verifier granted it.
In contrast to Bluesky accounts attested by trusted verifiers, a Blue
Badge is a way that ``any party may attach cryptographic attestations
to an AT Protocol record''~\cite{badge-blue}. More generally than
verified accounts granted by trusted verifiers, a Blue Badge may
indicate affiliation or group membership, but also attach specific
accomplishments or other recognition to an
identity~\cite{blue-badge-released}.

\subsection{Adversary Model}
\label{sec:adversary}

The general goal of our adversary is to disrupt users attempting
to access intended resources on the internet. This general goal
has two kinds of more specific goals, both of which we wish to counter.

We will say more about adversary capabilities after setting out
adversary goals in the next paragraph, but for simplicity in stating
those goals we sum these capabilities up in the
assumption that our adversary is able to disrupt any of the existing
mechanisms for certifying or validating (connection to) any meaningful
identity if the identity does not have a self-authenticating component.
Thus, users might intend to visit a given domain on the web or
interact with (read, follow, like, comment etc.) the data (posts,
comments, etc.)  of a given Bluesky handle. Our adversary is assumed
to be able to circumvent existing security mechanisms designed to
protect such access, for example, TLS\@.
accounts.

The main goal of our thus-capable adversary is to disrupt the discovery,
validation, or use of a self-authenticating meaningful identity. And the
primary instrumental goal to accomplish the main goal is to disrupt
discovery or validation of the association of a meaningful identifier
and a self-authenticating identifier. This may be simply in order to deny
service, but it may also be to hijack association.  Hijack may also be
simply to deny service, though less detectably so: a connection that appears
(misleadingly) to have succeeded in accessing an intended destination
is harder to detect than one that clearly fails---whether or not
failure includes an explicit error message of some kind.  But hijack
may be intended to support further attacks: surveilling behavior,
obtaining sensitive information from deceived clients, stealing client
credentials, etc.

Our adversaries are generally assumed to be able to perform DNS and TLS
hijacks (and BGP hijacks if needed~\cite{bgp-amazon,cert-hijack-hotpets17}).
Though there have been documented
cases of CA misbehavior, even if we assume all CAs are honest and follow
standards, adversaries can obtain TLS certificates for domains they
do not legitimately control~\cite{tls-hijack-ars,citp-blog-hijack}.
CT logs support the discovery of such
certificates, but only after the fact, so will not prevent misuse
before posting and discovery, and even then only if they are checked.
Further, this won't prevent typosquatting attacks or domain impersonation
tricks as in~\cite{domain-impersonation-acmccs2019}.
For some attacks that we consider, such a weaker adversary is sufficient to
carry them out.

In structural identifier association and certification, such as
employed in caching DNS records or in issuing TLS certificates, there
is a practical tension between requiring evidence of identity or of
control over a resource that is achievable by arbitrary honest
entities of modest capabilities and modest resources on the one hand,
and on the other hand, preventing adversaries from being able to
provide that same evidence for the identity or resources of an attack
target.  Despite substantial strides over the years, DNS and TLS
hijack remain a significant widescale threat.  Our intent is to
counter disruption even in the face of such an adversary.

Relatedly, Bluesky verified accounts as currently deployed might
resist our adversary, but that is in part because they are, by current
design, denied to the vast majority of Bluesky handles (currently
available to about .00017 of all handles). We will discuss in
Sec.~\ref{sec:contextual-trust} how the current verified accounts
system works and how we suggest making a system that is more broadly
usable while remaining secure. To be less succinct but more accurate,
we do not assume that the Bluesky verified account mechanisms as
currently designed and deployed can be circumvented by our
adversary. Rather, we are assuming an evolution of those mechanisms to
a more broadly available system for validating identities, one that
would generalize the existing system by resolving practical scaling
tensions in favor of broader general availability and correspondingly
more generally usable issuance validation checks for Bluesky handles
rather than in favor of meaningfully stronger identity assurance. This
would thus be if generalization were to follow roughly along the
trajectory that SSL/TLS certificates took as cryptographically
protected websites moved from being something specialized and rare to
ubiquitous and fundamental.  ``In the early days, CAs are strict about
identify verification before certificate issuance. Eventually, some
CAs realise that they can get away with less work and domain-validated
(DV) certificates are born.''~\cite{feisty-duck-history}

At the client end, our adversary has access to all
client traffic information that an ISP would have.  We are not directly
attempting by all of our techniques to counter adversaries blocking
access to the ``open'' internet in general, for example by blocking
access to VPNs. Nonetheless, some of our techniques make use of onion
services. Those will thus inherit all of the countermeasures to such
adversaries that using Tor clients provides, such as the possibility
of using obfuscated transport and nonpublic bridges or ephemeral proxies
for network access. 
Using onion services will also counter DNS manipulation and/or IP
blocking---assuming our techniques for secure association of an onion
address with a meaningful identifier are employed.

Tor clients use guards when connecting to the Tor network. The guard
is a node in the Tor network that the client select to use for all
connections through Tor. These were introduced to counter attacks
wherein an adversary induces repeated connections from the client
until the adversary controls both ends of a Tor connection and can
then perform end-to-end correlation~\cite{hs-attack06}.  Our adversary
is assumed to have control over a small fraction of Tor guards. The
exact size or upper bound on the fraction is not important for
purposes of this paper, but, to have a number to assist reader
comprehension, ten percent would be a reasonable ballpark assumption.
If the client is using Tor with a compromised guard, the adversary is
assumed to have access to all traffic information that a Tor entry
guard has. Though the adversary is assumed in this case to be able to
selectively block Tor streams or circuits, induce timing channels on
flows, etc., if the guard adversary simply blocks all Tor traffic or
is too disruptive, we assume the client will switch to another guard
or bridge that is not so disruptive.

\section{Synthesizing DID methods}
\label{sec:synthesizing-did-methods}

As~\cite{your-own-did-web} observes, a did-method-web DID gives the
identity owner greater control over their DID doc in the sense
that, ``If the plc.directory domain or future PLC resolution system
fundamentally changes or goes away, yours simply won't.'' On the other
hand, your DID doc is subject to the same blocking, hijacks (DNS,
TLS, etc.), and doppelganger attacks as is the domain on which it is
hosted. Also, updates to the DID doc of a did-method-plc identity
are recorded in a public append-only ledger, making them transparent.

Our first novel, though straightforward, observation is thus not the
application of onion association mechanisms to atproto identities.
Rather, using did-method-plc with a domain name you control as handle
is a simple way to have the advantages of both DID methods. We will
use the popular atproto event management site
{\small \texttt{smokesignal.events}} as an example.
The first step  was to list {\small \texttt{did:web:smokesignal.events}} under
{\small \texttt{alsoKnownAs}} in the DID doc
at {\small \texttt{https://plc.directory/did:plc:tgudj2fjm77pzkuawquqhsxm}}.
(Identifiers listed under {\small \texttt{alsoKnownAs}} in a DID doc are
considered to be identifiers of the same identity.)  Then we mirrored
primary fields from this DID doc using the {\small \texttt{did:web}} method
under
{\small \texttt{https://smokesignal.events/.well-known/did.json}}
and add in that DID doc\\
{\small \texttt{did:plc:tgudj2fjm77pzkuawquqhsxm}}
under {\small \texttt{alsoKnownAs}}. Now clients
subject to an adversary that blocks {\small \texttt{smokesignal.events}} but
not {\small \texttt{plc.directory}} (or vice versa) can still access fields
that are the same in the DID docs, particularly the key that
authenticates any records posted by either DID\@. The respective DID
docs reflecting these points are shown in subfigures (a) and (c)
of Figure~\ref{fig:did-docs}. (The inclusion of an onion address in
the respective {\small \texttt{alsoKnownAs}} lists, and the did-method-web DID
associated with that onion address in subfigure (b) will be discussed
in Sec.~\ref{sec:did-onion}.)

Clients that can access DID docs at both locations can check at
any time that fields are indeed the same (or how far the website is
behind the PLC in updates, if at all). An adversary that has hijacked
{\small \texttt{smokesignal.events}} for a subset of clients (e.g., based on
their IP range) could still offer a divergent DID doc in the
{\small \texttt{did.json}} file those clients see, which would only be
detectable by those with access to both the altered and unaltered
sites or the altered site and the PLC\@. Detection of malicious
behavior is important, but we can also preclude acceptance of the
bogus DID doc without the need to check such consistency. Clients
can be configured to only accept a DID association for
{\small \texttt{\@smokesignal.events}} that is attested to by endorsers they
trust. We will describe this in Sec.~\ref{sec:contextual-trust}.
Alternatively, if clients have a secure channel with someone they
trust on such matters, it is sufficient for them to receive on that
channel a single URL that incorporates both identifiers of the DID
association SAMI\@. This will be described in
Sec.~\ref{sec:dirt-simple-trust}.

\subsection{Handle hijack}
\label{sec:handle-hijack}

The atproto DID spec~\cite{atproto-spec-did} does not require a DID doc to include a handle ({\small
  \texttt{at://smokesignal.events}} in our example).  The
atproto handle spec~\cite{atproto-spec-handle} nonetheless
notes that handles should not be trusted as valid until the DID is
resolved and the current DID doc is confirmed to link back
to the handle. Placing the DID in a DNS TXT record is ``the
recommended and preferred resolution method for individual handle
configuration.''  Thus, an adversary capable of DNS hijack could break
bidirectional confirmation and hijack the handle if only the
{\small \texttt{did:web}} method is used.

Our entertained recommendation involves linking the
did-method-web DID to the did-method-plc DID by listing each under
{\small \texttt{alsoKnownAs}} in the other's DID doc. And the
{\small \texttt{did:plc}} Method Specification \emph{does} require that a
handle (an {\small \texttt{at://}} entry) be included under
{\small \texttt{alsoKnownAs}} in a did-method-plc DID
doc~\cite{atproto-spec-did-plc}. Of course an adversary could set
up its own did-method-plc DID that lists the handle
{\small \texttt{at://smokesignal.events}} under
{\small \texttt{alsoKnownAs}} in \emph{its} DID doc. Because the PLC
is a public append-only log, unlike with a DNS hijack, evidence of
such a DID doc will be guaranteed to appear in a permanent public
record. We will introduce in Sections~\ref{sec:did-onion}
and~\ref{sec:contextual-trust} further mechanisms so that connection
to such a hijacked domain handle will not only create public evidence,
but be detectable as part of connection exchanges. Those mechanisms
will require that, in the case of exactly two DIDs, one a
did-method-web DID and the other a did-method-plc DID, that their DID
docs must both list both DIDs as siblings (under either
{\small \texttt{id}} or
{\small \texttt{alsoKnownAs}} fields), and neither lists any other DID than
these two as siblings or under the {\small \texttt{controller}} field. Further,
the only handle listed is the one in the did-method-plc DID doc, which
must correspond to the did-method-web DID, e.g., {\small
  \texttt{at://smokesignal.events}} corresponding to {\small
  \texttt{did:web:smokesignal.events}}.
  
\subsection{DIDs in TLS Certificates}

Our next novel (if still straightforward) observation is that atproto
identities using did-method-plc can employ the subdomain SAN technique
described in Sec.~\ref{sec:basic-background-onion-association}. SAMIs
based on Bluesky handles that are subdomains of {\small
  \texttt{bsky.social}} are transparent via the PLC\@.  (Bluesky
supports handles as subdomains of {\small \texttt{bsky.social}} via
TLS-certificate wildcards rather than via explicit subdomain
inclusion. This makes account creation feasible for them, since
otherwise they would need a new certificate for each new handle. But
without explicit incorporation of such handles in TLS-certificate
SANs, the subdomain technique mentioned in Sec.~\ref{sec:related-work}
cannot work for those handles.)  When the handle in the did-method-plc
DID doc \emph{is} a DNS name other than a subdomain of {\small
  \texttt{bsky.social}}, then (as per our first observation) the
subdomain technique with the did-method-plc DID as subdomain of the
DNS name supports including that association in a TLS
certificate. Thus, for an atproto identity we can simply and
straightforwardly record the association of meaningful name and
self-authenticating name in both the PLC and CT logs.

We have applied techniques from secure onion
association to atproto identities without associated onion addresses. We
next turn to actually associating onion addresses with atproto identities.


\section{DIDs at .onions}
\label{sec:did-onion}

In our example, an adversary disrupting resolutions of or connections
to both {\small \texttt{plc.directory}} and {\small
  \texttt{smokesignal.events}} could disrupt DID resolution for the
handle {\small \texttt{@smokesignal.events}}. Also offering the DID doc
at an onion service makes it available at a much more
censorship-resistant location. The AT protocol explicitly does not
support onion addresses as handles~\cite{atproto-spec-handle}, but
nothing precludes hosting the DID doc at an onion address, which is
also a did-method-web subject, e.g., the \texttt{id} in the DID doc is
{\small
  \texttt{did:web:yrwoss6q2gxm5jpovzzubpr5i5y33jc2ru6wgpue}}
{\small
  \texttt{fviqf7rksnlfvaad.onion}}.

The onion address here is one set up to correspond to
{\small \texttt{smokesignal.events}} specifically as an
illustration of such DID doc hosting.
(Recall: Access to the onion service also assumes a client that speaks Tor
protocols and can route via the Tor network.)
To enforce mutual consent between this and an associated DID for a
registered-domain site, in this case, \texttt{did:web:smokesignal.events}, their
respective DID docs must reference each other.
This is illustrated by the three DID docs shown in
Figure~\ref{fig:did-docs}.

\begin{figure*}
\centering
\hspace*{-.1in}
\begin{subfigure}[b]{0.4\textwidth}
\tiny
\begin{verbatim}
{ "@context": [
     "https://www.w3.org/ns/did/v1",
     "https://w3id.org/security/multikey/v1"
      ],
  "id": "did:web:smokesignal.events",
  "alsoKnownAs": [
     "did:web:yrwoss6q  ---cut for spacing--- vaad.onion"
     "did:plc:tgudj2fjm77pzkuawquqhsxm"
      ],
  "verificationMethod": [
  {
     "id":                 "did:web:smokesignal.events#sign",
     "type":               "Multikey",
     "controller":         "did:web:smokesignal.events",
     "publicKeyMultibase": "zDnaeePHo ---cut for spacing---"
  }   ],
  "service": [
  {
     "id":                 "#smokesignal",
     "type":               "SmokeSignalService",
     "serviceEndpoint":    "https://smokesignal.events"
   }
 ]
}
\end{verbatim}
\label{fig:did-web-smokesignal}
\caption{Web DID document from\\ https://smokesignal.events/\mbox{.well-known}/did.json}
\end{subfigure}
\hspace{.6in}
\begin{subfigure}[b]{0.4\textwidth}
\tiny
\begin{verbatim}
{
  "@context": [
    "https://www.w3.org/ns/did/v1",
    "https://w3id.org/security/multikey/v1"
  ],
  "id": "did:web:yrwoss6q  ---cut for spacing--- vaad.onion",
  "controller": "did:web:smokesignal.events",
  "alsoKnownAs": [
    "did:plc:tgudj2fjm77pzkuawquqhsxm",
    "did:web:smokesignal.events"
  ],
  "verificationMethod": [
    {
      "id": "did:web:yrwoss6q---cut for spacing---.onion#sign",
      "type": "Multikey",
      "controller": "did:web:yrwoss6q ---cut for spacing---.onion",
      "publicKeyMultibase": "zDnaeePHo ---cut for spacing---"
    }
  ],
  "service": [
    {
      "id": "#attestation",
      "type": "DomainAttestationService",
      "serviceEndpoint": "https://yrwo ---cut for spacing---.onion"
    }
  ]
}
\end{verbatim}
\caption{Web DID document from\\ http://yrwoss6q2gxm5\ldots f7rksnlfvaad.onion/\mbox{.well-known}/did.json}
\end{subfigure}
\hfill
\newline
\newline
\begin{subfigure}[b]{0.3\textwidth}
\tiny
\begin{verbatim}
{ "@context": [
    "https://www.w3.org/ns/did/v1",
    "https://w3id.org/security/multikey/v1",
    "https://w3id.org/security/suites/ecdsa-2019/v1",
    "https://w3id.org/security/suites/secp256k1-2019/v1"
  ],
  "id": "did:plc:tgudj2fjm77pzkuawquqhsxm",
  "alsoKnownAs": [
    "at://smokesignal.events",
    "did:web:smokesignal.events",
     "did:web:yrwoss6q  ---cut for spacing--- vaad.onion"
  ],
  "verificationMethod": [
    {
      "id":                 "did:plc:tgudj2fjm77pzkuawquqhsxm#sign",
      "type":               "Multikey",
      "controller":         "did:plc:tgudj2fjm77pzkuawquqhsxm",
      "publicKeyMultibase": "zDnaeePHo ---cut for spacing---"
    }   

    {
      "id":                 "did:plc:tgudj2fjm77pzkuawquqhsxm#atproto",
      "type":               "Multikey",
      "controller":         "did:plc:tgudj2fjm77pzkuawquqhsxm",
      "publicKeyMultibase": "zQ3shgHhf ---cut for spacing---"
    }
  ],
  "service": [
    {
     "id":                 "#atproto_pds",
     "type":               "AtprotoPersonalDataServer",
     "serviceEndpoint":    "https://pds.cauda.cloud"
    }
  ]
}
\end{verbatim}
\caption{PLC DID document from\\ https://plc.directory/did/did:plc:tgudj2fjm77pzkuawquqhsxm}
\end{subfigure}
\caption{DID docs showing bi-directionality (mutual {\small \texttt{alsoKnownAs}} entries) and same signing key in each}
\label{fig:did-docs}
\end{figure*}

%

Besides {\small \texttt{alsoKnownAs}}, the W3C DID-core spec includes
a {\small \texttt{controller}}, which is not necessarily the same as the DID
subject. We use the {\small \texttt{controller}} property to create an
authorization graph where the onion DID explicitly delegates control
to the registered-domain identifier, establishing
that {\small \texttt{did:web:smokesignal.events}} possesses authority to
update the onion's DID doc. This verification method sharing across
DID docs (where the same {\small \texttt{publicKeyMultibase}}
appears in both DID docs with different {\small
  \texttt{controller}} designations), creates a cryptographic
authorization chain that proves either unified operation or explicit
delegation through private key sharing.

We assume an adversary capable of DNS and TLS hijack. It might thus
seem to be reducing the self-authenticating protection provided by
using an onion service to give
{\small \texttt{did:web:smokesignal.events}}
control over
{\small \texttt{did:web:yrwoss6q2gxm5jpovzzubpr5i5y33jc2ru6wgpuefviqf7rksnlfvaad.onion}}
(hereafter {\small \texttt{did:web:smoked.onion}} for
brevity). We should thus examine what is at risk. First and foremost,
the controller cannot change the id, and thus cannot change the key
that authenticates connection to {\small \texttt{http://smoked.onion}}.

If {\small \texttt{smokesignal.events}} were compromised, it could,
however, reset the key with which
{\small \texttt{did:web:smoked.onion}} signs records to a key that the
adversary knows but that the attack target does not. It could also add or
remove identities from {\small \texttt{alsoKnownAs}} in the DID
doc.  To do any of that, the DID doc updates would have to be
retrieved and applied. It is not automatic. If the maintainer of the
onion address, hence of {\small \texttt{did:web:smoked.onion}},
did not wish to accept any such update, it could simply not accept
it. And it can remove at any time any outside controller with which it
is dissatisfied.  The risk is not nonexistent, but it would require
distracted or reckless manual acceptance of problematic changes.

RFC 9421, HTTP Message Signatures, provides the active cryptographic
proof layer that transforms static DID document assertions into verifiable 
identifier trust~\cite{ietf-http-message-signatures-rfc}.
\cite{onion-discovery-attacks} describes onion-association attestation
headers that are checked by a browser WebExt.
Similarly, when the onion service signs its
HTTP responses using the {\small \texttt{Signature-Input}} and
{\small \texttt{Signature}} headers with a {\small \texttt{keyid}}
parameter explicitly
referencing {\small \texttt{did:web:example.com\#http-sig}},
it demonstrates real-time control of the keys declared in the DID docs. The
signature base construction includes critical message
components---{\small \texttt{@method}}, {\small \texttt{@authority}},
{\small \texttt{@path}},
{\small \texttt{content-digest}}, and {\small \texttt{date}}---following
RFC 9421 specs and signed with the Ed25519 algorithm.
A FLOSS reference implementation
of the WebExt and the server code, along with setup instructions, is
available. (See Sec.~\ref{sec:implementation}.)

Atproto operates on top of HTTPS\@. Thus, while blocking DID doc
access might be possible, access hijack also requires TLS hijack.  If
`hijack' is taken to include unicode substitutions, typosquatting, or
other forms of domain
impersonation~\cite{domain-impersonation-acmccs2019}, this is not
limited to just certificates for domains issued to other than those
who legitimately control them. In any case, connection hijack
involving TLS hijack \emph{sensu lato} continues to occur at
significant scale.
Onion association in DV TLS certs issued by any CA (e.g., Let's
Encrypt) can be accomplished by using the subdomain onion SAN approach
described above.  The just described technique using onion and
clearnet DIDs also accomplishes this without using the traditional PKI
infrastructure.

\subsection{Website names from atproto handles}
\label{sec:handles-for-website-names}

Prior work on onion association for SAMIs has been focused on
discovering onion addresses from DNS names and on doing so in a way
that is secure, transparent, blocking resistant, etc. But another
advantage that builds on the above is to allow you to have a website
with a meaningful name even when you do not control a registered
domain name.

A simple illustration can be contructed if you control the Bluesky
handle {\small \texttt{@example}} and the onion address {\small
  \texttt{[exampleonion].onion}}. In this case you can have their
respective DID documents point at and mirror each other as set out
above---just without also associating a DNS name at a third DID that
uses did-method-web with that DNS name. Discovery and security of
association are then as before. Without an associated clearnet URL,
access to the website assumes a client capable of accessing the Tor
network. But the most significant observation is that, if you can
obtain hosting for an onion service and can obtain a Bluesky account,
that is sufficient to have a website with a meaningful discoverable
name: you do not need to register a domain name.

There is also an efficiency and expediency advantage derived from
handle issuance. Assume you have already created an onion address and
put a lookup descriptor for it in the HSDir DHT\@. The steps involved
to obtain a standard Bluesky handle are few, and the time for the
did-method-plc DID doc (associating the handle with
the onion address) to be recorded in the PLC is on the order of
minutes. As described above, this is sufficient to have a publicly
available meaningful identifier that can be used to access a website.

Compare providing such an identifier by this means versus setting up
site access using a registered domain as meaningful identifier.  The
steps needed to register a domain name, configure DNS, and obtain a
TLS certificate are many more, and the time for all of those to be
processed is more on the order of hours. Using our decribed technique,
SAMI associations will at that point be tracked in the PLC\@. If you
were instead counting on CT logs alone rather than the PLC for
transparency, 24 hours is the typical merge delay---the maximum time a
CT log promises will pass before including a TLS certificate in a CT
log, as given in an SCT (signed certificate timestamp) that is
included in every TLS certificate.

Thus even if you ultimately want to have a handle that is a registered
domain name, if it is important to make the meaningful (and transparently
verifiable) identifier publicly available quickly and easily, setting
it up first as a standard Bluesky handle makes sense. And if you just
want a site discovery and access setup that is simpler and cheaper,
it also makes sense.

Simple ways to have a website with a meaningful name without needing
to register a domain name are also available via many other popular
services. For example, you can have a personal site if you have an
account at Github or Facebook. These sites are limited by the parent
service in the formats and content allowed, as well as limiting data
ownership, storage, and portability in ways incompatible with basic
stated goals of atproto for user control, ownership, and portability
of their own data.  Using the atproto handle and onion service as
described, however, your website can offer anything that an onion
service can.

An atproto handle must be a DNS name. The proper handle from our
illustration is {\small \texttt{example.bsky.social}}.  Another
important distinction of this approach versus traditional sites such
as Github and Facebook, however, is that you are relying on Bluesky
\emph{only to supply a meaningful name and associated link to a
service address}. In that case, it might appear that the handle could
be used to just point at a webserver at an IP address, rather than an
onion address. But accessing this server would not be verified by
a self-authenticating identifier, amongst other problems.
By having the website at an onion address, the lookup and
association obtains the mutual authentication of our SAMI design. 

At time of writing, Bluesky is the only atproto handle issuer for
which DID docs are tracked in public, append-only logs. Thus for now,
this is a way in which the {\small \texttt{did:plc}} method is not
yet meeting the atproto goals of decentralized ownership and
portability. There are plans in motion both to provide PLC mirrors and
to move the PLC out of direct control of
Bluesky~\cite{independent-PLC}. But for now, you need a Bluesky handle
and an onion address to straightforwardly create an atproto SAMI
website with transparency and resistance to hijack, blocking,
fingerprinting, etc., if you do not control a registered domain.

It would be useful if the atproto handle supported recognition by the
browser (or initially, a WebExt) which would then automatically connect
to the onion service. For example, the extension could
detect a handle with a particular format such as {\small
  \texttt{@example-onyn}}. Until {\small \texttt{bsky.social}} can be
expected to serve the appropriate header, this could be included in a
pinned post or similar that would imply a lookup and connection to the
associated onion service. Or a standard site supporting such
association could be set up (similar to the API offered at {\small
  \texttt{sauteed-onions.org}} for discovering and validating onion
association via CT logs). Further, onion services using \emph{client
authorization} support additional lookup and privacy protections for
onion services~\cite{client-authorization}, which could complement rapidly
developing ATmosphere work on private data and permissioned spaces.
(See, for example, {\small \texttt{https://ngerakines.leaflet.pub/}}
  for blog entries about ``permissioned data''.)
To support uniqueness and portability of names in keeping with the
spirit of atproto, account providers could agree to only issue account
handles with such specialized format in a manor that supports such
global uniqueness and portability.

No standardization by atproto account providers nor any existing
WebExt as yet supports any of the above. And we expect to pursue these
in future work. Nonetheless, as set out in the ``simple illustration''
above, it is already possible to support manual discovery of an onion
address website from an atproto handle, which will then be validated by
the signatures and DID doc once connected to that website. This can be
done manually now without any cooperation by, or even awareness of,
Bluesky.

\subsection{Dirt simple trust}
\label{sec:dirt-simple-trust}

Receiving a web address from a trusted source over a trusted channel,
e.g., via a Signal message from someone you know and trust, is not
generally sufficient by itself to achieve SAMI.
\emph{Dirt simple trust} occurs when learning an address from a
trusted source is sufficient for a client to connect to that address
with assurance of SAMI~\cite{secdev19}.
We have created a modified version of the WebExt described
in~\cite{onion-discovery-attacks} that validates
dirt simple trust, but based on the DID rather than an onion address.
(Below, in Section~\ref{sec:implementation}, we provide a brief
description of the current implementation, along with a pointer to the
code, instructions, and documentation.)
The dirt-simple SAMI URL for
our ongoing example is
{\small \texttt{https://smokesignal.events/?check=}}
       {\small \texttt{did:plc:tgudj2fjm77pzkuawquqhsxm}},
and the DID doc
checked by the browser extension does not need to
include an onion address. Our WebExt for Firefox and Tor Browser
validates association of meaningful and self-authenticating
identifiers when connecting to such a SAMI\@.  With either approach to
dirt simple trust, consistency and transparency are provided by the
PLC, but if the association is also included via
subdomain in the TLS cert as described above in
Sec.~\ref{sec:synthesizing-did-methods}, then consistency between
the PLC and CT logs can also be checked.

Dirt simple trust using DID association offers greater simplicity.
But dirt simple trust using onion association provides greater
blocking and manipulation resistance by automatically routing to the onion
address in the path of the dirt-simple SAMI\@.
Such rerouting was implemented in the previously mentioned WebExt
for SATAs. We intend, but have not yet implemented this functionality
in our WebExt under development.
%

In most instances, association via secure communication from a source
trusted to assert such association is not available.  If the
association (DID or onion) is in the TLS certificate, a transparent
and censorship resistant means of discovery is provided by CT logs.
Censorship resistance is offered simply by having discovery afforded
by CT logs or log monitors such as \texttt{crt.sh}. But offering a log
monitor itself at an onion service provides even greater blocking and
manipulation protection while preserving
transparency~\cite{sauteed-onions-wpes}.

Absent dirt simple trust, someone in control of an
onion address can manipulate control over a domain with no more
resistance than otherwise offered against TLS hijack. Such vulnerability
also applies to DID association with a domain. There are
structural solutions just as there are for showing domain ownership
during TLS cert issuance. Amongst their limitations, they are at odds
with the general spirit of decentralized identities. We next turn
to this and to the alternative of providing contextual roots of trust.


\section{Contextual Trust}
\label{sec:contextual-trust}

In late 02024, Alexios Mantzarlis ``found that 44 percent of Bluesky's
100 most-followed accounts have a
doppelganger''~\cite{engadget-on-bluesky-impersonation-problem}.
As observed by Mantzarlis at the
time~\cite{bluesky-impersonation-problem}, Bluesky had already
recommended and given instructions for how a website owner could
verify their Bluesky account by setting their web address as
handle~\cite{bluesky-website-handle-instruction}.  But as we noted and
cited in previous sections, typical website discovery and
authentication are subject to various forms of hijack. Further, one
does not need to own or even just DNS hijack \texttt{example.com} to
obtain the handle {\small \texttt{@example.bsky.social}}. And doppelgangers
Mantzarlis uncovered were often simply typosquats and similar. For
example, one of the more prominent doppelganger handles (since
suspended) was {\small \texttt{@officialmarkhammil.bsky.social}}.

As noted in Sec.~\ref{sec:handles-for-website-names}, the Bluesky
default handle format permits something like domain
squatting.  The official atproto handle of the Tor Project is {\small
  \texttt{@torproject.org}}. When writing this paper we 
encountered {\small \texttt{@torproject.bsky.social}} and thought it
might be another handle set up by someone at the Tor
Project. Personal communication indicated otherwise, and that account
has since been suspended.
In part to combat such problems, in April 02025 Bluesky introduced
\emph{verified accounts} as attested by a \emph{trusted verifier}.
``For example, the New York Times can now issue verification badges to
its journalists directly in the app.''~\cite{bluesky-trusted-verifier-announced}.  Verified
accounts display a ``user-friendly, easily recognizable verification
badge'' on their {\small \texttt{bsky.app}} profile, a blue circle with a white
check mark. Trusted verifiers display a similar badge but with scalloped
edges.

According to numbers at time of writing from
{\small \texttt{https://bverified.vercel.app/}} there are 7835 verified
accounts across 20 trusted verifiers. This amounts to .00017 of
Bluesky's 45 million accounts. But this is
not just because verified accounts are only about a year old and
have not yet taken off. There is an intrinsic tension between
``verified'' being a broadly available designation and it 
providing meaningful assurance. In fact, there are two
such tensions. One was
already noted in Sec.~\ref{sec:adversary}: if control over a
meaningful identifier is something an account holder of modest resources and
capabilities can generically prove to a general verifier, it is hard to
prevent attackers with more significant resources and capabilities
from being able to offer such proof for the same identifier, or
one easily mistaken for it (e.g., involves
a typo or unicode substitution, etc.).

The other tension is probably more intractable at scale: 6694 of
the verified accounts (85\%) are verified by Bluesky with the
remaining verifications occurring across the other nineteen trusted
verifiers, all but a few of which are newsmedia or journalism sites or
organizations. To be verified by those others assumes an
affiliation with the verifier. Without affiliation, Bluesky
requires that ``[y]ou, or your account, must be notable within your
field and geographic region,'' according to the official Bluesky
Verification Request Form on Google
Docs~\cite{bluesky-trusted-verifier-announced,bluesky-verified-account-req-form}. To
the extent this policy is a response to the problem reported on
by Mantzarlis, it makes sense. Most people would like to know that
Bluesky activity by Mark Hamill is activity of the famous actor, not
anyone else.  If all of us could be verified in a way that somehow
overcame the tension noted in the previous paragraph, having a
verified account badge would nonetheless cease to provide assurance
beyond that our account is genuinely our account.  Without a
presumption of prominence, some assurance of context is needed for
there to be any meaning to the assurance that an identity is ``who you
think it is''. Verified account badges are insufficient to provide
this.  (Though not exactly Zooko's
Triangle~\cite{zookos-triangle}, it seems to be
closely related.)

The current system was a response that could be rolled out fairly
quickly to a pressing problem for handles including those of Bluesky's
most prominent, well-known users. According to personal communication
with Bluesky personnel, the longer term intent is to provide a system
that works for a much larger fraction of the userbase.

\subsection{Background on onion association SATAs with contextual trust}

Contrast Bluesky verified accounts with the attestation for onion
association mentioned in the
previous section. (Because such attestation is
by a SATA attestating about itself or another SATA,
\cite{onion-discovery-attacks} calls it a \emph{sattestation}.  The
term is also applicable to the broader Self-Authenticating Meaningful Identity of this paper.)
In addition to self-sattestations already described in
Sec.~\ref{sec:did-onion}, there are third-party-SAMI
\emph{sattestors} issuing sattestations. These are signed by the onion
key of the sattestor and attest to the binding of the domain name and
onion address of a sattestee. There can also be a contextual label,
and labels can also support delegation.  Following an example
in~\cite{satlog-csf}, a sattestor with label for U.S.\ Govt.,
{\small \texttt{sattestor(USG)}},
can issue sattestations to a SAMI with the {\small \texttt{USG}} label or
with ``narrower'' labels {\small \texttt{USG:DOJ}}
or~{\small \texttt{USG:DOJ:FBI}}.

Trust in SAMI sattestors and sattestor labels is also contextual. A NY
Times SAMI should not be trusted to make sattestations about arbitrary
identities, only those with a \texttt{nytimes} label.  Such contextual
trust can be doubly important given the increasing number of
journalists who are not employees of organizations for which they are
writing. And a local community site, such as for a chamber of
commerce, can sattest that a given merchant or media site is local to
that area. This can counter widescale international misinformation
operations masquerading as local news sites to leverage greater trust
therein~\cite{paperwall}.

Sattestations and labels as described
in~\cite{secdev19,onion-discovery-attacks} were implemented for a
server frontend and for extensions for Firefox and Tor Browser, but
these implementations have not been updated in many years. Our
implementation intended to offer similar and additional functionality
for atproto handles and DIDs is available and in active development.

\subsection{Contextual trust for Bluesky trusted verifiers}
\label{sec:contextual-trust-for-trusted-verifiers}

Though accounts verified by Bluesky itself have no context (other than
the context of being ``notable''), Bluesky trusted verifiers do offer
an element of meaningful contextual relevance: the verifier of a
verified account is displayed if one clicks on the verified account
badge.  But in the system of~\cite{onion-discovery-attacks}, a
third-party (SATA) sattestation is not trusted by clients unless the
sattestee attests itself to have the associated identifiers and the
label---another way sattestations better reflect the atproto
spirit of autonomy than do trusted verifications. Relatedly, such
sattestations facilitate \emph{authority-independent authentication}:
no authority can usurp control over an identity from its
owner~\cite{secdev19}.  And even if one is unconcerned about autonomy
at this point, if the number of trusted verifiers grows, so does the
chance that some verifier ``takes credit'' for activity of an account
by assigning it a contextual label associated with the verifier, a
label that the assignee does not accept.

Our Firefox and Tor Browser extension does not yet have implemented
the check for agreement with the sattestation by a sattestor of local
(self-sattestation) labels indicated by a record signed by a key in
the DID doc of the SAMI\@. This could, once implemented, serve as a
more general and autonomy-respecting replacement for the current
version of Bluesky trusted verifiers. If rooted in (delegatable)
sattestation from Bluesky, this would fully replicate and generalize
the existing system. Even without that, however, it is possible to
integrate the existing Bluesky trusted verifier system with the
greater autonomy of sattestations in a backwards compatible way.

The SATA sattestations of~\cite{onion-discovery-attacks} assume that
trust roots are part of the local client configuration, much like
browsers are configured to trust a standard set of CAs for TLS
certificates but are configurable to add to (remove from) that
list. Assuming wider acceptance of our approach, one can assume some
standard trust roots will be preconfigured or available as initial
configuration options, somewhat like existing CAs.  We are working on
expanding our FF/TB extension to usably support default and
configurable atproto SAMI sattestor trust roots. If one then took
Bluesky trusted verifier status as extrinsic evidence that one should
trust a sattestor, then e.g., \texttt{(@nytimes.com,
  did:plc:eclio37ymobqex2ncko63h4r)} could be configured as locally
trusted to have the label \texttt{sattestor(nytimes.com)}.  So, any
browser with our extension expanded to include this feature could be
configured to trust such sattestations from Bluesky trusted verifiers
even if Bluesky does not itself adopt the use of sattestations.

\subsection{Integration with AT Protocol}
\label{sec:integation-with-atproto}

We have applied a contextual trust approach to atproto identities, 
where integration with atproto's identity infrastructure suggests 
a broader vision for compositionally constructed contextual trust 
relationships. DID docs can incorporate verifiable presentations 
for proving organization membership or community endorsement, with HTTP 
signatures binding credentials to active service provision.

If Alice wants to endorse Bob as a journalist in the town of Springfield,
she can leverage DID infrastructure in two ways. For \emph{inline 
sattestation}, Alice creates a signed Content Identifier (CID) of her 
endorsement content, which Bob then stores in his PDS with both 
signatures. The endorsement is bound through the DID 
verification methods, with HTTP signatures providing real-time proof of 
key possession as content is served.
For \emph{remote sattestation}, Alice posts the CID in her repository at 
her PDS, where it is picked up by the firehose. Bob 
receives notification and, if he accepts, puts an acceptance in his 
repository including a pointer to her proof. This approach allows Alice 
to revoke an endorsement since it remains entirely under her control,
similar to how the {\small \texttt{controller}} property
in DID docs can be updated to revoke delegated authority.

Basic remote-sattestation mechanisms rely on the atproto relay, which
currently does not speak to Tor. A censor who can block Alice's PDS
can effectively block her endorsement of Bob for anyone unable to
reach it. For inline sattestation with DID-based verification,
immediate prospects are better---if Bob mirrors his repository at an
onion service (identified {\small \texttt{serviceEndpoint}} in his DID
  doc), then anyone reaching that onion service can still
  access Bob's records, including Alice's endorsement, even if their
  access to the firehose is blocked.

This creates a trust framework where context is not merely implicit in
validation procedures but explicitly declared, cryptographically
verifiable, and composable across multiple trust domains---enabling
progressive enhancement where basic clients just follow {\small
  \texttt{serviceEndpoint}}s while more manipulation-resilient clients
perform full cryptographic verification including credential
validation.


\section{Implementation Notes}
\label{sec:implementation}

The reference implementation for the functionality described
throughout this paper, along with instructions for creating a test
server, for setting up a server, and for WebExt installation and
setup, can be found at
{\small \texttt{https://github.com/ngerakines/domain-attestations/}}.
The implementation demonstrates signing and verification of
DID docs using widely adopted tools rather than
custom cryptographic solutions, unlike the implementations
of prior SATA approaches~\cite{secdev19,onion-discovery-attacks}, which predate
RFC 9421. Two interchangeable server-side signers are
provided: server-py, a lightweight Python Flask application that generates a
{\small \texttt{did:web}} document from an ECDSA P-256 key and signs responses
to
{\small \texttt{/.well-known/did.json}}; and {\small \texttt{caddy-httpsig}},
a Go middleware plugin that applies RFC 9421 signatures to any
response from an existing Caddy site. Both signers generate standard
{\small \texttt{Signature}} and {\small \texttt{Signature-Input}}
headers using ecdsa-p256-sha256, with a keyid that
references a {\small \texttt{verificationMethod}} within the signed
document. Additionally, a companion Firefox WebExt intercepts
responses via the webRequest API, retrieves the referenced DID
doc, and verifies the signature against the embedded public key
using the browser's native Web Crypto API.

As noted in Sec.~\ref{sec:synthesizing-did-methods},
the DID spec states, ``It is crucial to validate the
handle bidirectionally, by resolving the handle to a DID and checking
that it matches the current DID document.'' And in the handle
spec~\cite{atproto-spec-handle}, ``Handles should not be
trusted or considered valid until the DID is also resolved and the
current DID document is confirmed to link back to the handle. The link
between handle and DID must be confirmed bidirectionally, otherwise
anybody could create handle aliases for third-party accounts.''
Despite bidirectionality being essential to valid association
of a handle with a DID, it appears that users are
expected to do this manually---for something so crucial, a surprising
expectation of both concern and technical competence on the part of all 45
million Bluesky users, along with all the other atproto users.  There
is currently no straightforward usable mechanism for checking such
validation of which we are aware. Precisely such a validation check
is, however, among those that our WebExt provides.
Though not for all handles, it does the validation check for any
handle that is also the name of a domain being visited. We believe
ours is the first available tool for doing so.


\section{Conclusion}
\label{sec:conclusion}

We have set out application of existing means of secure onion
association to atproto identities and means to combine both approaches
in a single identity. We have also described the advantages such
combining engenders, such as stronger and more broadly available
blocking resistance and manipulation resistance both for the
association of identifiers in identities and for identity-associated
content. Looking forward, we will be enthused to support more central
adoption by the ATmosphere of our more decentralized approach to
verified identities.



\section*{Acknowledgments}
  We would like to thank Rasmus Dahlberg, Aaron Jaggard, and a few
  anonymous commentators for helpful comments and suggestions on
  drafts of this paper.

\bibliographystyle{plain}
\bibliography{onionloc.bib}

\end{document}